\documentclass[]{pasj01}

\Received{}
\Accepted{}
 
\usepackage{lscape}
 
\begin{document} 

\title{ 
Spin evolution of neutron stars in wind-fed high mass X-ray binaries
}

\author{Shigeyuki \textsc{karino}\altaffilmark{1}%
}
\altaffiltext{1}{Kyushu Sangyo University, 2-3-1 Matsukadai, Higashi-ku, Fukuoka 813-8503, Japan}
\email{karino@ip.kyusan-u.ac.jp}


\maketitle

\begin{abstract}

The observed X-ray pulse period of OB-type high-mass X-ray binary (HMXB) pulsars are typically longer than 100 seconds. 
It is considered that the interaction between the strong magnetic field of neutron star and the wind matter could cause such a long pulse period.

In this study, we follow the spin evolution of NS, taking into account the interaction between the magnetic field and wind matter.
In this line, as new challenges, we solve the evolution of the magnetic field of the neutron star at the same time, and additionally we focus on the effects of wind properties of the donor. 
As the result, evolutionary tracks were obtained in which the neutron star spends some duration in the ejector phase after birth, then rapidly spins down, becomes quasi-equilibrium, and gradually spins up.
Such evolution is similar to previous studies, but we found that its dominant physics depends on the velocity of the donor wind.
When the wind velocity is fast, the spin-down occurs due to magnetic inhibition, while the classical propeller effect and settling accretion shell causes rapid spin-down in the slow wind accretion. 
Since the wind velocity of the donor could depend on the irradiated X-ray luminosity, the spin evolution track of the neutron star in wind-fed HMXB could be more complicated than considered.

\end{abstract}

\section{Introduction}

A high-mass X-ray binary (HMXB) is a binary system in which mass accretion occurs from a massive donor to a relatively young compact object, such as a neutron star.
Because the neutron star in a HMXB is bright and has a strong magnetic field, it is an ideal experimental site for extreme physics such as neutron star matter, general relativity, strong magnetic field, and so on \citep{ST82}.
HMXBs are classified into roughly two groups according to their donor type: OB-type and Be-type \citep{C84,C86,B97}. 
In a Be-type HMXB, mass transfer from a circumstellar disk around a Be donor to a neutron star occurs. 
On the other hand, a neutron star in OB-type system accretes matter overflown from the Roche lobe of the OB donor, or accrete from the strong stellar wind from the donor. 
Furthermore, the OB-type systems could be categorized into several sub-groups such as persistent sources, super-giant fast X-ray transients (SFXTs), and highly obscured systems \citep{W03,S05,Z05,M09,CR12}.

It is known that in Be-type systems, there is a strong correlation between the spin period of neutron stars and the orbital period of the systems \citep{C84,C86,V20}.
On the other hand, the neutron stars in wind-fed HMXBs show a broad distribution of spins from several seconds to several hours \citep{MN17}.
Unlike Be-type systems, in wind-fed systems, there are no clear correlation between neutron star spin and the orbital period of the binary systems \citep{C84,C86}.

It is broadly considered that the spin of the neutron star is related to their magnetic field. 
A neutron star is considered to be born with a rapid rotation and a strong magnetic field.
When this strong magnetic field interacts with the wind matter, the spin angular momentum is extracted and the neutron star spins down rapidly.
After the rapid spin-down, the neutron star spin gradually evolves under the balance 
between the magneto-rotational centrifugal force and the accretion pressure.
Although the magnetic field plays critical role on the spin of the neutron stars, however, the magnetic fields of the neutron stars measured by cyclotron resonant scattering feature (CRSF) in HMXBs show almost the constant degree around $B \sim 10^{12} \rm{G}$ \citep{C17,T19}. 
In addition, howemver, as another point of view, the fact that the observed HMXB magnetic field is $10^{12} \rm{G}$ or more may be due to an observational bias.
If the magnetic field is much weaker than this level, the accretion column cannot be bound so strongly that the X-ray emission region is widened, which may make it difficult to detect X-ray pulsations.
On the contrary, when the magnetic field is too strong the electron-cyclotron energy shifts into high energy range and becomes difficult to detect in the CRSF observation.
Additionally, it has been pointed out that if a much stronger field is obtained, it may be difficult to distinguish from the ion-cyclotron feature \citep{S02}.
Even if there are observational biases, the broad distribution of the spin of neutron stars with similar magnetic fields means that there are other factors that affect spin besides the magnetic field.

In recent years, on the other hand, studies on linking magnetic field evolution and spin evolution of neutron stars have been advanced \citep{PT12,D16,WT20}.
Especially, around accreting neutron star from the stellar wind, wind matter should be decelerated by a shock formation and then the settling accretion matter forms thermally supported shell \citep{S12}.
Due to the shell formation, wind matter cannot be accreted to the neutron star straightforwardly.
In such a shell,  
the angular momentum carried by the wind matter will be inevitably smoothed.
As the result, the accreting matter does not simply bring the angular momentum onto the neutron star, but the spin variation of the neutron star depends on the interaction of the shell matter and the magnetic field line of the neutron star \citep{S12,S17}.

\citet{PT12} first tried to explain the spin property of the slowly rotating 
neutron star in HMXB by applying the accretion shell model.
They considered that the accretion shell is formed around a young neutron star, and compute the spin evolution of the neutron star, at the same time with the magnetic field evolution. 
Then they showed that the observed spin and spin-down rate can be reconstruct under such an evolution scenario.
This result suggests that at least in some HMXB systems, the shell formation could play crucial roles on the spin evolution of neutron stars.

Most of previous studies, however, only consider relatively slow stellar winds implicitly or explicitly because it is considered that 
X-ray emission from neutron stars hinders the stellar wind acceleration \citep{S91,K12}.
Inversely, however, in relatively dim phase of neutron star X-ray, wind acceleration in the donor could work effectively.
Hence, in this study, in order to reconstruct a distribution of neutron star spin in wind-fed HMXBs, a series of computations of the spin and magnetic field evolution, focusing on the wind velocity. 
As the result, we find that the wind velocity is important on the dominant physical processes in the spin evolution of the wind-fed neutron star.

This paper is organized as follows: in the next section we describe our method to obtain the spin evolution of neutron stars.
In section 3, we show the obtained results.
Especially, we show the variability of spin evolution track depending on the wind feature. 
In section 4, we compare our results with observations and discuss on our model in detail.
In the last section we present a conclusion.

\section{Binary evolution model}

In this study, we consider a binary system which consists of an OB-type donor and an accreting neutron star.
In such a system, the neutron star emits bright X-ray due to the mass accretion from the stellar wind of the donor. 
Hereafter, we fix the initial mass and radius of the neutron star as the typical values: $M_{\rm{NS}} = 1.4 M_{\odot}$ and $R_{\rm{NS}} = 1.0 \times 10^{6} \rm{cm}$, respectively. 
On the other hand, we choose the donor parameters as follows: the mass, $M_{\rm{d}} = 14.9 \, \rm{M}_{\odot}$, the radius, $R_{\rm{d}} = 6.3 \, \rm{R}_{\odot}$ and the luminosity, $L_{\rm{d}} = 2.5 \times 10^{4} \rm{L}_{\odot}$, respectively.
This donor model corresponds to the initially $15 \, \rm{M}_{\odot}$ single star at the age of 5 Myr, computed by the approximated stellar evolution track \citep{H00}. 
In a HMXB, the accreting neutron star was born from a supernova explosion of the initial primary star. 
Assuming that the initial primary was $20 \rm{M}_{\odot}$ at zero-age, the life-time of this initial primary is about 5 Myr, and hence we adopt this value as the present age of the donor. 
The main sequence life-time of the $15 \, \rm{M}_{\odot}$ star is about 10 Myr. 
Hence, the modeled donor has not expanded, and has not filled Roche lobe yet.
Since the time-scale of the spin evolution of the accreting neutron star is much shorter,
we do not consider the evolution of the donor (except for the mass) and lock in the donor parameters in the computations.

The orbital separation is obtained from masses of binary components and orbital period. 
Since the donor star in such a system loses its mass via wind mass loss (and some of the wind matter are captured by the neutron star), the binary orbit changes continuously.
Also in a tight system, tidal interaction and gravitational wave emission will change the orbital motion gradually.
We take these orbital evolutions into account, following the formulae given in \citet{H02}.
In this study, we set the initial orbital period and the eccentricity as $P_{\rm{orb}} = 5 \rm{d}$ and $e = 0.1$, respectively.
As a reference, the observed examples of wind-fed systems are listed in Table~\ref{tab:1}.

The mass loss rate of the donor, depending on the mass, radius and luminosity, is described by the recipe in \citet{V01}.
Then due to the line-driven mechanism (CAK mechanism), the matter will be accelerated \citep{CAK75}.
The matter accretion is considered to be proceeded via Bondy-Hoyle-Littleton (BHL) theory \citep{B44,HL39}.
If the matter falls onto the neutron star, it brings in some amount of angular momentum, and hence the neutron star spins up.
On the other hand, the interaction of the rotating magnetic field with the wind matter can reduce the spin rate via the so-called propeller effect \citep{IS75,DP81}.
Also neutron star spins down when the angular momentum is pulled off due to magnetic dipole radiation emission.

Recently, it has been shown that a settling accretion shell supported by thermal pressure will be formed around a wind-fed neutron star \citep{S12}.
The interaction between this settling shell and the magnetic field of the neutron star could dominate its spin evolution, and hence we need to take such a shell formation into account.

Regardless of the working physical processes, the spin down mechanisms strongly depend on the strength of the magnetic field of the neutron star.
Hence, the decay of the magnetic field of the neutron star affects the spin evolution of the neutron star significantly.
Therefore, it is critically important to solve the evolution of the magnetic field when we consider the spin evolution of the neutron star.

\begin{table}
	\centering
	\tbl{Examples of HMXB pulsars \footnotemark[$*$].
	}{
	\begin{tabular}{lcc}
		\hline
		object & $P_{\rm{spin}} [s] $ & $P_{\rm{orb}}$ [d]     \\
		\hline
		OAO1657-415 & 37.7 & 10.4 \\
		IGR J18027-2016 & 139.8 & 4.6 \\
		Vela X-1 & 283 & 8.96 \\
		4U1907+09 & 439 & 8.37 \\
		4U1538-52 & 529 & 3.73 \\
		IGR J16393-4643 & 904 & 3.7 \\
		SXP1062 & 1062 & 300 ? \\
		IGR J16493-4348 & 1069 & 6.78 \\		
		2S0114+650 & 9700 & 11.6 \\		
		\hline
	\end{tabular}}
\begin{tabnote}
\footnotemark[$*$] Data is taken from \citet{T11,PT12,MN17}.  
\end{tabnote}
\label{tab:1}
\end{table}

\subsection{Magnetic-field evolution}

It is considered that the neutron star has strong magnetic field just after its birth, because of the compression of the field lines of its progenitor and the strong convection deep inside of the supernova explosion \citep{TD93,S08,N20}. 
This initial strong field declines mainly due to Ohmic diffusion and Hall diffusion in earlier stage of neutron star evolution. 
To involve this magnetic field evolution, we adopt the approximated evolution model of the neutron star magnetic field: 
\begin{equation}
\label{eq:B-field}
B(t) = \frac{ B_{\rm{ini}} \exp (-t / \tau_{\rm{O}} ) }
{ 1 + (\tau_{\rm{O}} / \tau_{\rm{H}}) \left[ 1 - \exp  (-t / \tau_{\rm{O}} ) \right] } 
+ B_{\rm{fin}}
\end{equation}
\citep{A08,PT12,K19}.
In this recipe, we assume the time-scale of Ohmic diffusion $\tau_{\rm{O}}$ and Hall diffusion $\tau_{\rm{H}}$ as $\tau_{\rm{O}}= 1 \rm{Myr}$，and $\tau_{\rm{H}} = 1 \rm{kyr}$, respectively.
Here, we need two more parameters: initial field $B_{\rm{ini}}$ and final (relic) field $B_{\rm{fin}}$.
We set the final field as $B_{\rm{fin}} = 10^{9} \rm{G}$, which is the typical value observed in neutron stars in old systems such as low-mass X-ray binaries (LMXBs).

\begin{figure}
 \begin{center}
  \includegraphics[width=8cm]{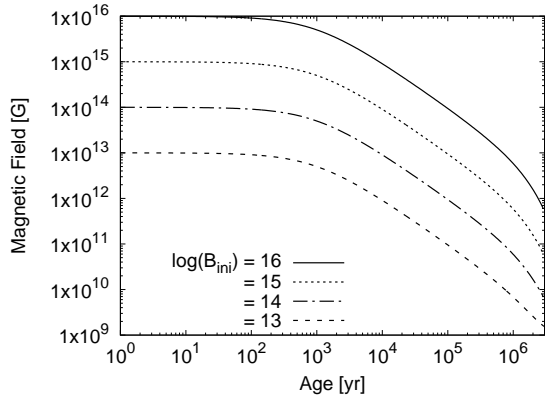} 
 \end{center}
\caption{
The evolutions of the magnetic field are shown for several values of initial field: $B_{\rm{ini}} = 10^{13} - 10^{16} \rm{G}$.
The relic field is fixed as $B_{\rm{fin}} = 10^{9} \rm{G}$.
}
\label{fig:B-field}
\end{figure}

The initial strength of magnetic field is still under debate.
In this study, we vary this initial value as a parameter, and we examine $\log(B_{\rm{ini}}) = 14$, 15 and 16.
This range is chosen since the observed neutron stars in HMXBs, which age should be $\sim 1 \rm{Myr}$, have typically similar magnetic field around $10^{12} \rm{G}$ or more.    
In Fig.~\ref{fig:B-field}, the time evolutions of the magnetic field for each initial conditions are shown. 
The magnetic field decrease drastically after $\sim 1 \rm{kyr}$, and this field decay brings a large difference from the simplified computation of the spin evolution with the constant field strength \citep{D16,K19,WT20}. 
In the simulation of a supernova explosion focused on the initial magnetic field of a proto-neutron star, a dynamo produces a magnetic field of up to $10^{15} \rm{G}$.
And since only weak supernova explosions can be reproduced by numerical calculation, it is not surprising that stronger convection in actual supernova produces a stronger magnetic field than the current simulated predictions \citep{N20}.

When the neutron star accretes matter, the field magnetic field is embedded in the neutron star surface and the field strength declines \citep{ZK06}. 
In old systems such as LMXBs, this effect could be important mechanism of magnetic field evolution. 
In wind-fed systems, however, since accretion rate is small and accretion time-scale is short, accretion induced field decay may not play an important role.

\subsection{Wind accretion model}

The wind matter from the donor is accelerated by line-driven mechanism, which is also known as CAK mechanism \citep{CAK75}. 
In this study, we compute wind velocity by the standard line-driven model: 
\begin{equation}
v_{\rm{w}} = v_{\infty} \left( 1 - \frac{R_{\rm{d}}}{r} \right)^{\beta} .
\label{eq:wind}
\end{equation}
$\beta$ is the parameter governing the degree of acceleration; small $\beta$ gives strong acceleration and wind matter quickly approaches to the terminal velocity $v_{\infty}$.

According to \citet{V01}, the terminal velocity of the massive star becomes $k \times v_{\rm{esc}}$, 
where $v_{\rm{esc}} = \left( 2 G M_{\rm{d}} R_{\rm{d}}^{-1}  \right)^{1/2}$ denotes the escape velocity.
The numerical factor $k$ is 2.6 for hot stars and 1.3 for less hot stars. 
Since the surface temperature of the present stellar model is higher than the critical temperature, the terminal velocity of the wind 
for the present donor model becomes $2.5 \times 10^{8} \rm{cm \, s}^{-1}$.

\begin{figure}
 \begin{center}
  \includegraphics[width=8cm]{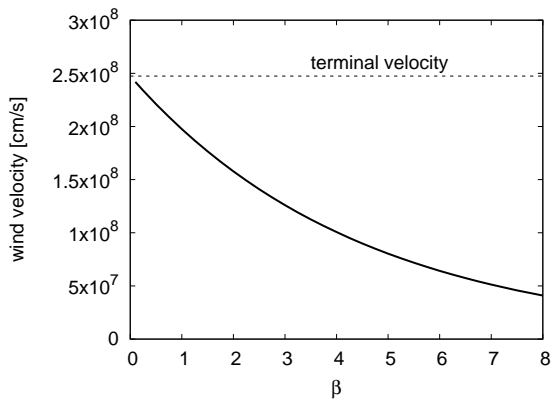} 
 \end{center}
\caption{The wind velocity at the neutron star position, as the function of the parameter $\beta$ in our model.
}\label{fig:wind}
\end{figure}

In the present study, we show that the wind velocity of the donor ($v_{\rm{w}}$) plays an important role on the spin evolution of the neutron star.
To see the importance of the wind velocity, we set the terminal velocity as the value given above and examine several values of parameter $\beta$. 
Namely, we control the wind velocity at the position of the accreting neutron star by changing $\beta$; $\beta = 1$ (fast wind) to $\beta = 7$ (slow wind).
The wind velocity at the neutron star position as a function of $\beta$-parameter is shown in Fig.~\ref{fig:wind}.

In the binary evolution code given by \citet{H02}, the wind velocity is artificially controlled as to be much smaller than the escape velocity. 
Also in the actual observations of HMXBs, the wind velocities of some
donors seem to be smaller than those of single OB-type stars \citep{KKK15,GG16,H20}. 
For example, the wind velocity of the well-studied Vela X-1 is estimated as only $7 \times 10^{7} \rm{cm \, s}^{-1}$ \citep{GG16}.
These slow wind could be caused by the X-ray irradiation from the neutron stars in these systems. 
The efficiency of the line-acceleration mechanism strongly depends on the ionization factor of the outer envelope of the donor \citep{S91,K12}, and we revisit this topic in Section 4.

The mass loss rate from the donor is computed by the formulae given in \citet{V01}.
In their model, the mass loss rate depends on the donor mass, luminosity, temperature, metallicity, and the terminal velocity of the wind. 
The metallicity is fixed to be $Z = 0.02$ and other parameters are set to be the values introduced above. 
With our donor model, the resultant mass loss rate due to the wind becomes $\dot{M}_{\rm{w}} = 4.8 \times 10^{-8} \rm{M}_{\odot} \rm{yr}^{-1}$ .

From the obtained wind parameters ($v_{\rm{w}}$ and $\dot{M}_{\rm{w}}$), the mass accretion rate onto the neutron star is obtained. 
In this computation, the standard BHL accretion model is used \citep{B44,HL39}. 
In BHL theory, wind matter is accreted when it comes into the accretion radius ($R_{\rm{acc}}$), where the potential of the neutron star gravity becomes larger than the wind kinetic energy: 
%
\begin{equation}
R_{\rm{acc}} = \frac{2GM_{\rm{NS}}}{v_{\rm{w}}^{2} } . 
\label{eq:Racc}
\end{equation}
Then the mass accretion rate onto the neutron star is given as 
\begin{equation}
\dot{M}_{\rm{acc}} = \frac{R_{\rm{acc}}^{2}}{4 a^{2}} \dot{M}_{\rm{d}} ,
\end{equation}
where $a$ denotes the semi-major axis of the binary orbit.
When the orbit is not circular, we need to multiply a factor $\left( 1 – e^{2} \right)^{-1/2}$ to this mass accretion rate in order to average it over the orbit 
\citep{BJ88,H02}.

According to the results of numerical simulations of fast-wind accretions, flip-flop instability may disturb mass accretion \citep{M88, BP11}. 
In this case, the mass accretion rate could be frequently fluctuated.  
Moreover, in tight binary systems, it is suggested that the wind stream could be focused in \citep{HC12,EM19a}. It is too complicated to treat these phenomena, and therefore we limit our study in the range of the standard BHL theory.

\subsection{Spin evolution mechanisms and their torque} 

It is considered that a newly born neutron star has strong magnetic field and spins rapidly.
The break-up limit of the neutron star spin period is less than 1ms. 
However, since many kinds of instabilities could make the spin slower, the spin period of the nascent neutron star could be longer \citep{C70,FS78,K00,SK04}.
Hence, in this study, we assume the initial period of the neutron star as 10 ms.
From such a nascent neutron star, strong electro-magnetic radiation and relativistic particles are emitted and mass accretion onto the neutron star could be avoided; this evolutionally stage is called the {\it{ejector phase}} \citep{PT12,D16}.

When the out-going flux is larger than the accretion ram pressure, the mass accretion is inhibited. 
By using mass conservation law and the standard pulsar flux, the critical spin period is
\begin{equation}
P_{\rm{ej}} = 2 \pi \left( \frac{4 \mu^{2} }{3 \dot{M}_{\rm{acc}} v_{\rm{w}} c^{4} } \right)^{1/4} ,
\label{eq:Pej}
\end{equation}
where we simply assume $\mu = B R_{\rm{NS}}^{3}$ \citep{PT12}. 
Here we consider that the neutron star is in the ejector phase, when 
\begin{equation}
P_{\rm{ej}} > P_{\rm{spin}}.
\label{eq:eje}
\end{equation}
Note that the dividing point between ejector phase and propeller phase is the point where ram pressure and out-going flux are balanced at the light cylinder radius or gravitational radius.
In the present study, this balance is evaluated at $R_{\rm{acc}}$, because it comes outside the the light cylinder, except when very fast wind accretes on a slowly rotating neutron star.

On the other hand, during the ejector phase, the spin-down due to electro-magnetic wave radiation emission continuously proceeds. 
Assuming a dipole radiation, the spin-down torque is given as
\begin{equation}
N_{\rm{dp}} = \frac{2}{3} \frac{\mu^{2} }{c^{3}} \left( \frac{2 \pi}{P_{\rm{spin}}} \right)^{3} 
\end{equation}
\citep{G07}.

After the end of the ejector phase, the wind matter coming inside of the accretion radius $R_{\rm{acc}}$ is captured by the neutron star gravity. 
Then, at the magnetic radius, where the magnetic pressure balances to the accretion ram pressure, the accreting matter is trapped by the magnetic field lines of the neutron star.
If, at this magnetic radius, the rotation rate of the magnetic field lines is larger than the Kepler rotation rate, the centrifugal force dominates and matter will be blown away; this process is known as the propeller effect \citep{IS75}. 
The condition that the propeller effect sets in is 
\begin{equation}
R_{\rm{cor}} < R_{\rm{mag}} < R_{\rm{acc}}. 
\end{equation}
Here, the corotation radius $R_{\rm{cor}}$ and magnetic radius $R_{\rm{mag}}$, where the magnetic energy density is equal to the kinetic energy density of the accretion matter, is
\begin{equation}
R_{\rm{cor}} = \left( \frac{G M_{\rm{NS}} P_{\rm{spin}}^{2}}{4 \pi^{2}} \right)^{1/3}
\end{equation}
and 
\begin{equation}
R_{\rm{mag}} = \left( \frac{\mu^{2} a^{2}}{2 \dot{M}_{\rm{d}} v_{\rm{w}} } \right)^{1/6} ,
\label{eq:Rmag}
\end{equation}
respectively \citep{G07,B08}. 
In the propeller phase, the spin angular momentum is transferred to the ejected wind matter through contact with the rotating magnetic field.
Consequently, the neutron star spin declines rapidly. 
This spin-down torque is
\begin{equation}
N_{\rm{prop}} = \left( \frac{1}{36} \frac{\mu^{4}}{G M_{\rm{NS}}} \frac{ 4 \pi^{2}}{P_{\rm{spin}}^{2} R_{\rm{mag}}^{3}} \right)^{1/2} 
\label{eq:nprop}
\end{equation}
\citep{G07}.
After the rapid spin-down due to the propeller effect, the spin of the neutron star settles in the quasi-equilibrium rate, 
\begin{equation}
P_{\rm{eq}} = \left( \frac{8 \pi^{4} \mu^{2} a^{2}}{G^{2} M_{\rm{NS}}^{2} \dot{M}_{\rm{w}} v_{\rm{w}} } \right)^{1/4}
\label{eq:prop}
\end{equation}
and the spin rate gradually increases with the decline of the magnetic field strength, as described in Eq.~(\ref{eq:B-field}).

On the other hand, in the situation of a quasi-spherical wind accretion, it is suggested that a thick settling accretion shell could be formed around the neutron star \citep{S12}. 
This shell is cooled mainly due to the inverse-Compton mechanism and matter in the shell gradually falls toward the neutron star with the decline of the thermal pressure. 
When the matter density is low, the cooling time-scale in the shell could be longer than the convection time-scale of the shell matter. 
According to Shakura et al. (2012), if the condition 
\begin{equation}
\dot{M}_{\rm{acc}} < \dot{M}_{\rm{crit}} \approx 4 \times 10^{16} \rm{g \, s}^{-1}
\label{eq:mdcrit}
\end{equation}
is satisfied, settling accretion shell will inhibit direct accretion. 
In this case, convection in the shell could contribute to averaging the angular momentum brought by the accretion matter.
Settling shell matter finally interacts with the magnetic field lines of the neutron star, and through this interaction, the angular momentum will be exchanged between the neutron star and the shell matter. 
The direction of the transport of the angular momentum depends on the spin rate and magnetic field strength.
The torque acts on the neutron star is described as follows:
\begin{equation}
\label{eq:shell}
N_{\rm{shell}} = A \dot{M}_{\rm{acc}}^{7/11} – B \dot{M}_{\rm{acc}}^{3/11} ,
\end{equation}
where $A$ and $B$ are the functions as follows:
\begin{eqnarray}
A &=& 4.6 \times 10^{31} K \mu_{30}^{1/11} v_{8}^{-4} P_{\rm{orb, 10}}^{-1},  \\
B &=& 5.5 \times 10^{32} K \mu_{30}^{13/11} P_{\rm{spin,100}}^{-1}, 
\end{eqnarray}
in cgs unit \citep{LSL16,PT12,S17}.
$\mu_{30}$ is the magnetic dipole moment in the unit of $10^{30} \rm{G \, cm}^{3}$.
$P_{\rm{orb, 10}}$ and $P_{\rm{spin, 100}}$ denotes the orbital period and the spin period normalized by 10 days and 100 seconds, respectively. 
The constant $K$ is a non-dimensional value and here we set this value as $K \approx 40$ \citep{PT12}, though the coefficients of these equations varies slightly in each previous studies.

Roughly speaking, the spin evolution when the settling shell is formed is as follows. 
The rapidly rotating neutron stars loses its spin angular momentum due to the interaction with the accretion shell. 
Hence, the neutron star quickly spins down. 
Once the spin of the neutron star reaches a kind of quasi-equilibrium state (the right hand side of Eq.~(\ref{eq:shell}) becomes zero ), the spin rate changes gradually with the decay of the magnetic field strength.

Wind matter may interact with the magnetic field lines before they are captured by neutron star gravity.
In this case, {\it{magnetic inhibition}} in which the spin angular momentum of the neutron star is directly extracted by the wind matter occurs.
We assume that the accretion material and the magnetic field collide elastically.
The spin-down torque in this stage is based on the assumption that the stellar wind matter is swung at the rotational speed of the neutron star at the magnetic radius, and evaluate by
\begin{equation}
N_{\rm{mi}} = R_{\rm{mag}} \left( \dot{M} v_{\rm{spin}} \right) .
\end{equation}
Here, $\dot{M}$ is the mass entering into the magnetic radius within the unit time, and $v_{\rm{spin}}$ denotes the rotational velocity of the magnetic field lines at the magnetic radius.

\section{Spin evolution of neutron stars and its dominant processes}

Applying the method introduced above, we compute the time evolution of the spin of the neutron star. 
We start the time evolution from the birth of neutron star, and compute the neutron star spin and magnetic field.
To trace the time-developing, we set the time-step as $1/100$ of the neutron star age.
However, when the spin varies drastically, we modify the time-step so that the change of the spin angular momentum should be less than 5 \% in one time-stepping.
At the same time, we also compute the orbital parameters of the system. 
Because of small mass transfer, however, the orbit varies only slightly. 
Hence, hereafter, we do not mention the orbital change in this study.

Here we vary the initial magnetic field of the neutron star from $10^{14} \rm{G}$ to $10^{16} \rm{G}$ and obtain the spin evolution for each cases. 
We assume the CAK wind model; the $\beta$ parameter in Eq.~(\ref{eq:wind}) is a free parameter and we vary this $\beta$ from $\beta = 1$ (fast wind) to $\beta = 7$ (slow wind). 
See also Fig.~\ref{fig:wind}.

Regardless of the choice of the parameter sets, the neutron star stays in ejector phase for a certain duration after its birth.
During this phase, the neutron star loses its angular momentum due to dipole radiation and gradually spins down. 
After a period of time, when the spin becomes slow and the magnetic field decays, 
the spin evolution of the neutron star proceeds to the next phase.

After the ejector phase ends, 
the spin evolution of the neutron star depends on the size relationship between 3 radii: magnetic radius $R_{\rm{mag}}$, accretion radius $R_{\rm{acc}}$ and corotation radius $R_{\rm{cor}}$. 
If the accretion radius is larger than the magnetic radius, the wind matter will be captured gravitationally by the neutron star. 
Furthermore, if the magnetic radius is smaller than the corotation radius, the wind matter can be accreted further inside in the accretion radius.
If the accretion rate is smaller than the critical value (see Eq.~(\ref{eq:mdcrit})), however, a settling accretion shell will dominate the spin evolution of the neutron star via Eq.~(\ref{eq:shell}) \citep{S12}. 
Inversely, if the accretion rate is enough high, wind matter could be accreted onto the neutron star.
On the other hand, if the magnetic radius is larger than the corotation radius, the wind matter captured by the neutron star will be blown away; the propeller spin-down, shown in Eq.~(\ref{eq:nprop}), starts \citep{IS75,DP81,B08}.

At the end of the ejector phase, if the magnetic field is still strong and the magnetic radius is larger than the accretion radius, the wind matter will encounter the magnetic field lines before it is captured by neutron star gravity. 
Especially, when the magnetic radius is larger also than the corotation radius, the wind matter interacted by the field lines will be blown off by a strong centrifugal force. 
This situation corresponds to the super-Keplerian magnetic inhibition phase discussed in \citet{B08}.
During this phase, the neutron star spins down rapidly and its corotation radius becomes larger.
Once the corotation radius overcomes the magnetic radius, the wind matter captured by the field lines will not be blown away and form a shock at a certain position. 
Then they will fall onto the neutron star within the time-scale of the Kelvin–Helmholtz instability \citep{HL92}. 
This phase corresponds to the sub-Keplerian magnetic inhibition phase \citep{B08}. 
In this phase, the quasi-equilibrium spin period of the neutron star decrease gradually due to the decay of the magnetic field. 
Since this quasi-equilibrium state achieves under the balance between the magneto-centrifugal force and the accretion pressure, 
the neutron star takes the same track of the spin evolution of that due to the propeller effect, described in Eq.~(\ref{eq:prop}).

Some examples of the obtained spin evolution tracks are shown in Figs.~\ref{fig:fast} -- \ref{fig:slow}. 
In Fig.~\ref{fig:fast}, the cases with the wind parameter $\beta = 1$ (fast wind) are shown for different values of initial magnetic field strength: (a) $B_{\rm{ini}} = 10^{14}\rm{G}$，(b) $10^{15}\rm{G}$ and (c) $10^{16}\rm{G}$, respectively.
In a similar way the result for $\beta = 4$ (intermediate wind) and $\beta = 7$ (slow wind) are shown in Fig.~\ref{fig:intermediate} and Fig.~\ref{fig:slow}.

At first, we consider Fig.~\ref{fig:fast} (fast wind case) in detail.
Here we are assuming that the orbital period of the system is 5 days, then the wind velocity at the neutron stellar position becomes $2.0 \times 10^{8} \rm{cm \, s}^{-1}$. 
In this case, within several thousand years, the magnetic field declines significantly 
and the ejector phase finishes.
Since the wind velocity is fast, the accretion radius is smaller than the magnetic radius at this point. 
On the other hand, the spin of the neutron star is still rapid, the rotation rate becomes super-Keplerian at the magnetic radius. 
As the result, a rapid spin-down due to the super-Keplerian magnetic inhibition occurs just after the end of the ejector phase.
And then the neutron star comes in a quasi-equilibrium state. 
In this stage, the neutron star spins up gradually with the decay of the magnetic field. 
When we change $B_{\rm{ini}}$, the starting time of the spin-down and the spin period during the quasi-equilibrium state become altered.
Namely, when the magnetic field is weak, the ejection phase becomes shorter and the spin period after the rapid spin-down becomes longer.

The spin evolutions of the neutron star accreting from the wind with intermediate velocity ($1.0 \times 10^{8} \rm{cm \, s}^{-1}$ at the neutron star orbit) are shown in Fig.~\ref{fig:intermediate}.
Their early evolutions are similar to the case with fast wind accretion. 
In this case, however, after the decay of the magnetic field, the accretion radius overcomes the magnetic radius. 
Then, the wind matter will be trapped by the gravitational field of the neutron star and fall further inside. 
During this process, the wind matter will go through a shock and forms a subsonic settling accretion shell. 
In this phase, the interaction of the shell and the magnetic field dominates the spin evolution of the neutron star, as discussed in previous studies \citep{S12,PT12,S17}.
The neutron star spins down rapidly again, when this shell-dominant phase starts. 
After achieving the quasi-equilibrium state, the spin evolves along this quasi-equilibrium line with keeping the right-hand-side of Eq.~(\ref{eq:shell}) zero.
After the slow spin-up, the corotation radius comes inside of the magnetic radius, and then the propeller effect becomes active. 
Hence, the further evolution of the spin follows the quasi-equilibrium evolution due to the magneto-centrifugal balance.

The early spin evolutions for slow wind cases ($5.1 \times 10^{7} \rm{cm \, s}^{-1}$; Fig.~\ref{fig:slow}) seem also similar to those shown in above qualitatively. 
In these cases, however, the accretion radius has already become larger than the magnetic radius, at the end point of the ejector phase. 
Hence, the rapid spin down after the ejector stage occurs due to the classical propeller effect \citep{IS75,DP81}. 
On the other hand, when the initial magnetic field is strong, since the magnetic radius is still large, the spin evolution becomes similar to the intermediate wind cases shown in Fig.~\ref{fig:intermediate}.

\begin{landscape}

\begin{figure*}
  \includegraphics[width=6cm]{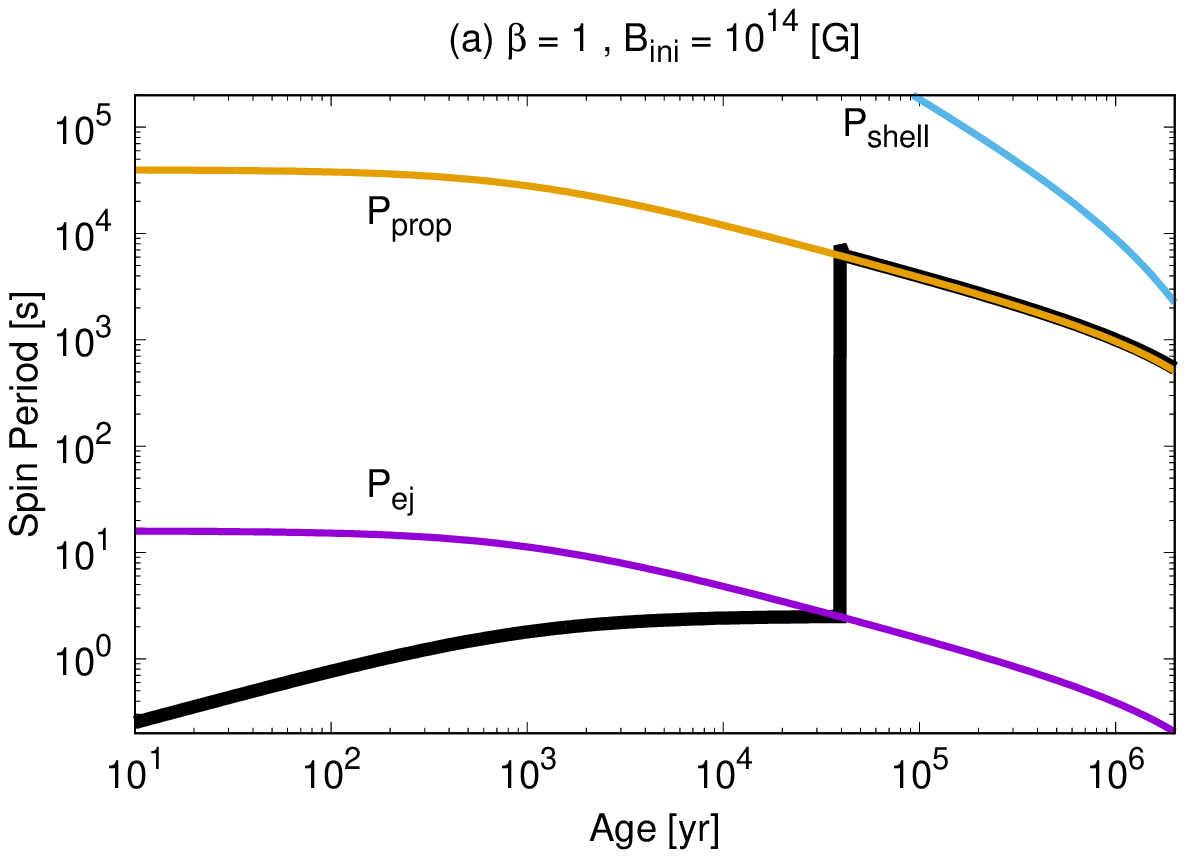} 
  \includegraphics[width=6cm]{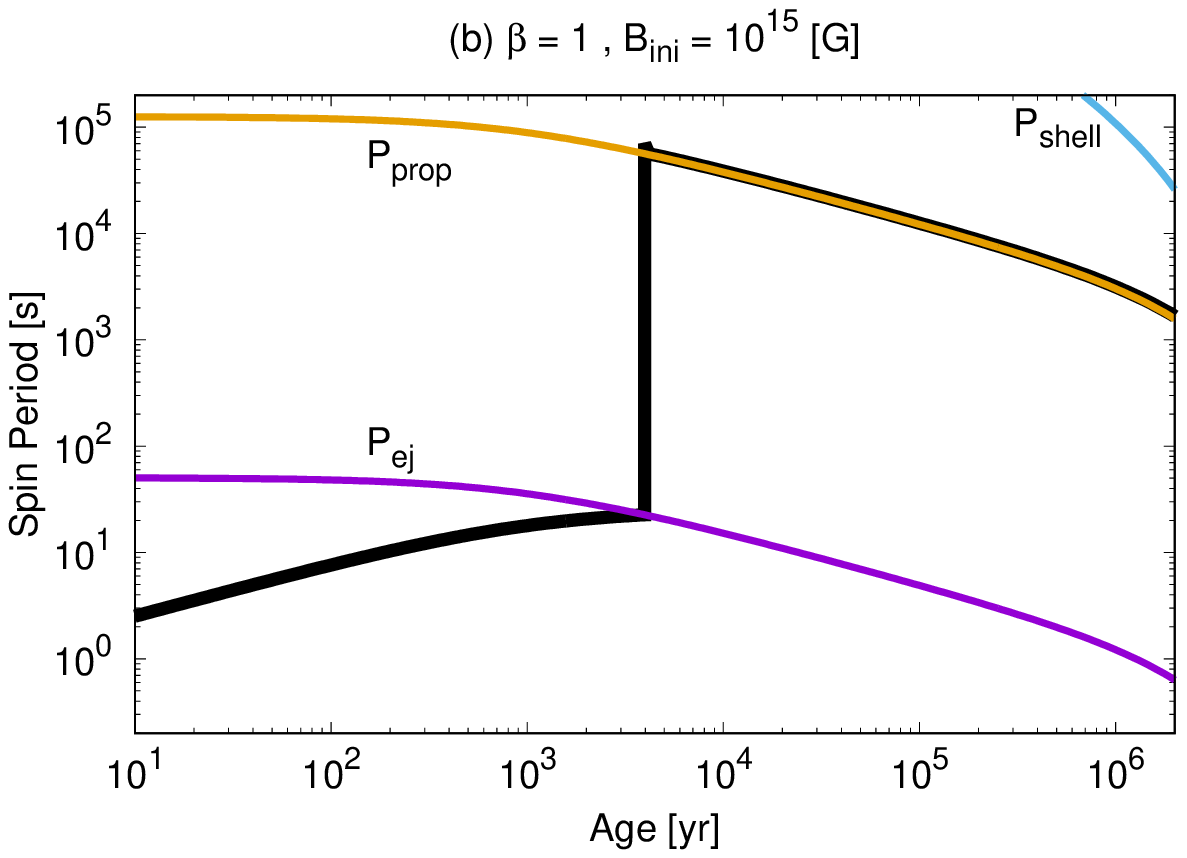} 
  \includegraphics[width=6cm]{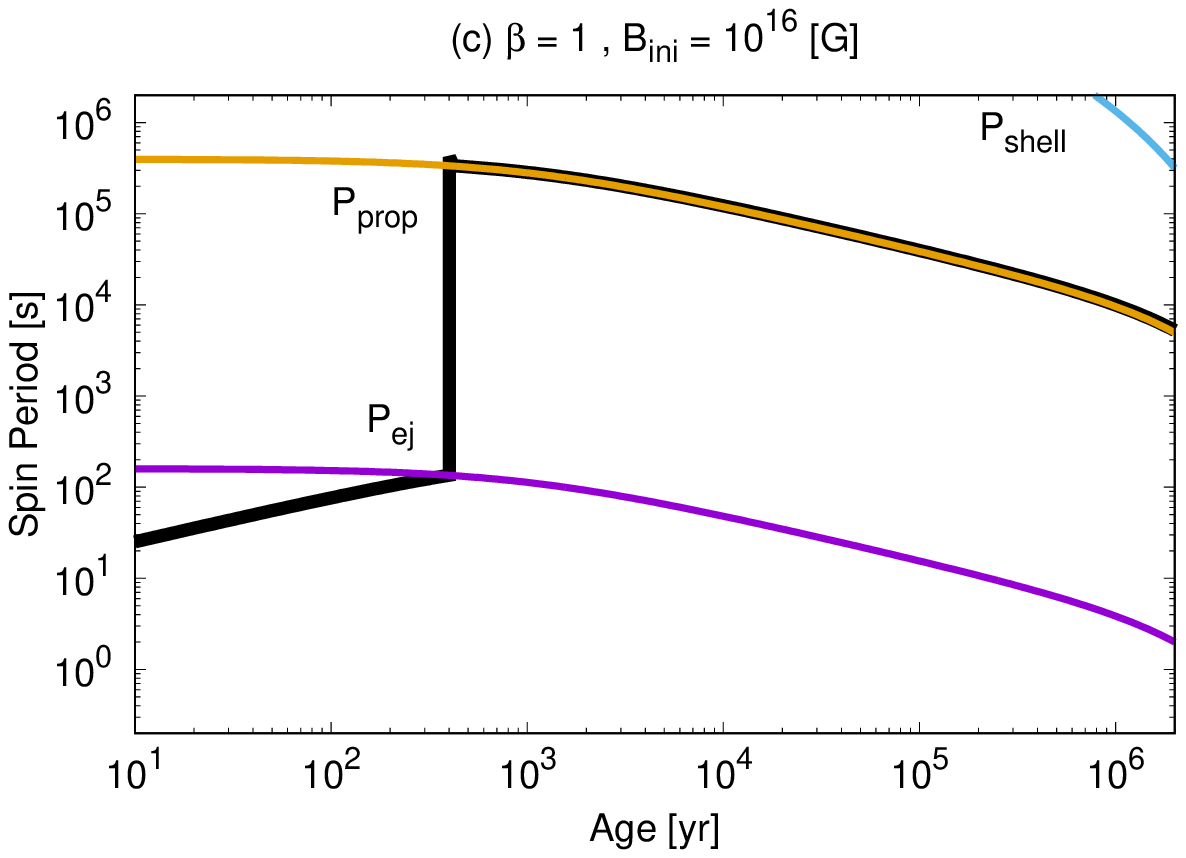}
\caption{
The spin evolution of the neutron star in wind-fed HMXB. 
The black solid lines denote the computed spin evolutions of the neutron star, while thin curves denote three typical periods: 
ejector condition $P_{\rm{ej}}$ with magenta line, period corresponding to the spin equilibrium  $P_{\rm{prop}}$ with orange line, 
and the quasi-equilibrium state due to shell interaction $P_{\rm{shell}}$ with blue line, respectively.
The wind velocity parameter is set as $\beta = 1$, that is, the wind velocity is fast.
The three different initial fields are examined: (a) $B_{\rm{ini}} = 10^{14} \rm{G}$, (b) $B_{\rm{ini}} = 10^{15} \rm{G}$
and (c)$B_{\rm{ini}} = 10^{16} \rm{G}$.
}
\label{fig:fast}
\end{figure*}

\begin{figure*}
  \includegraphics[width=6cm]{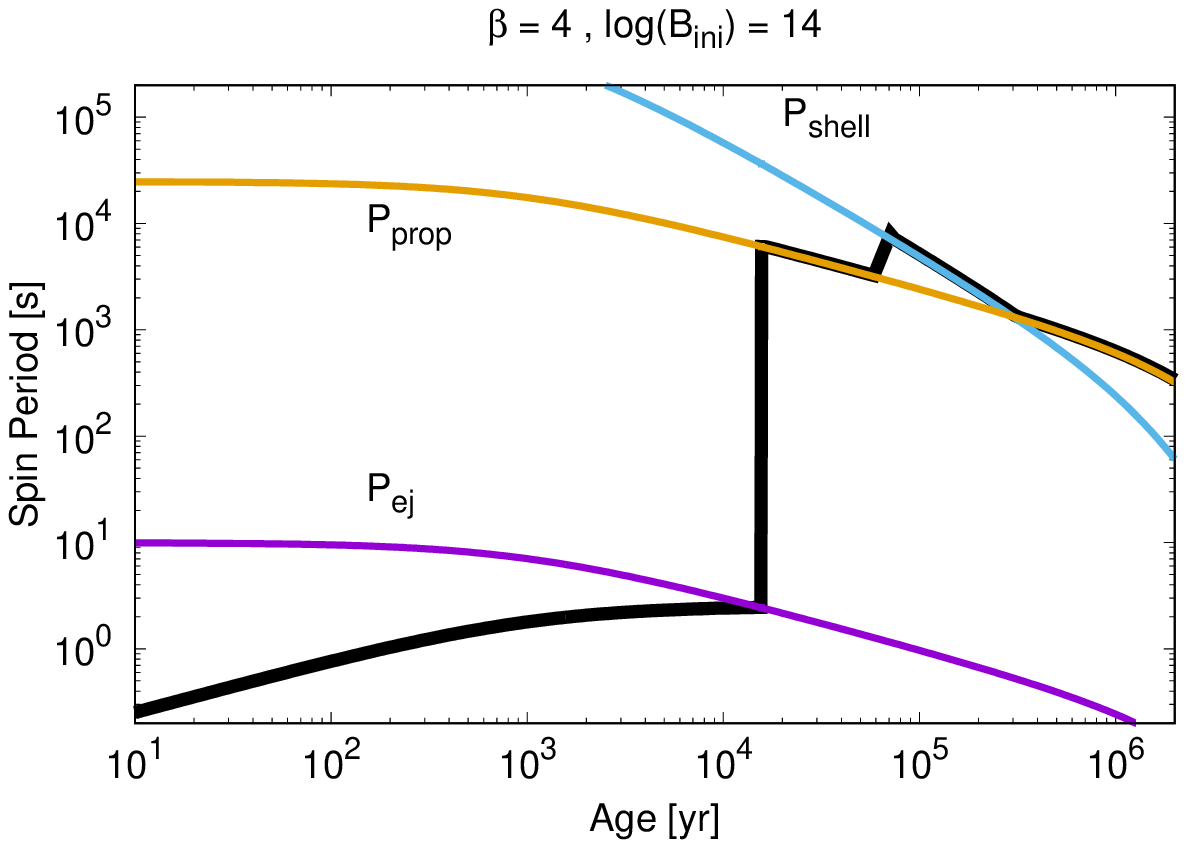} 
  \includegraphics[width=6cm]{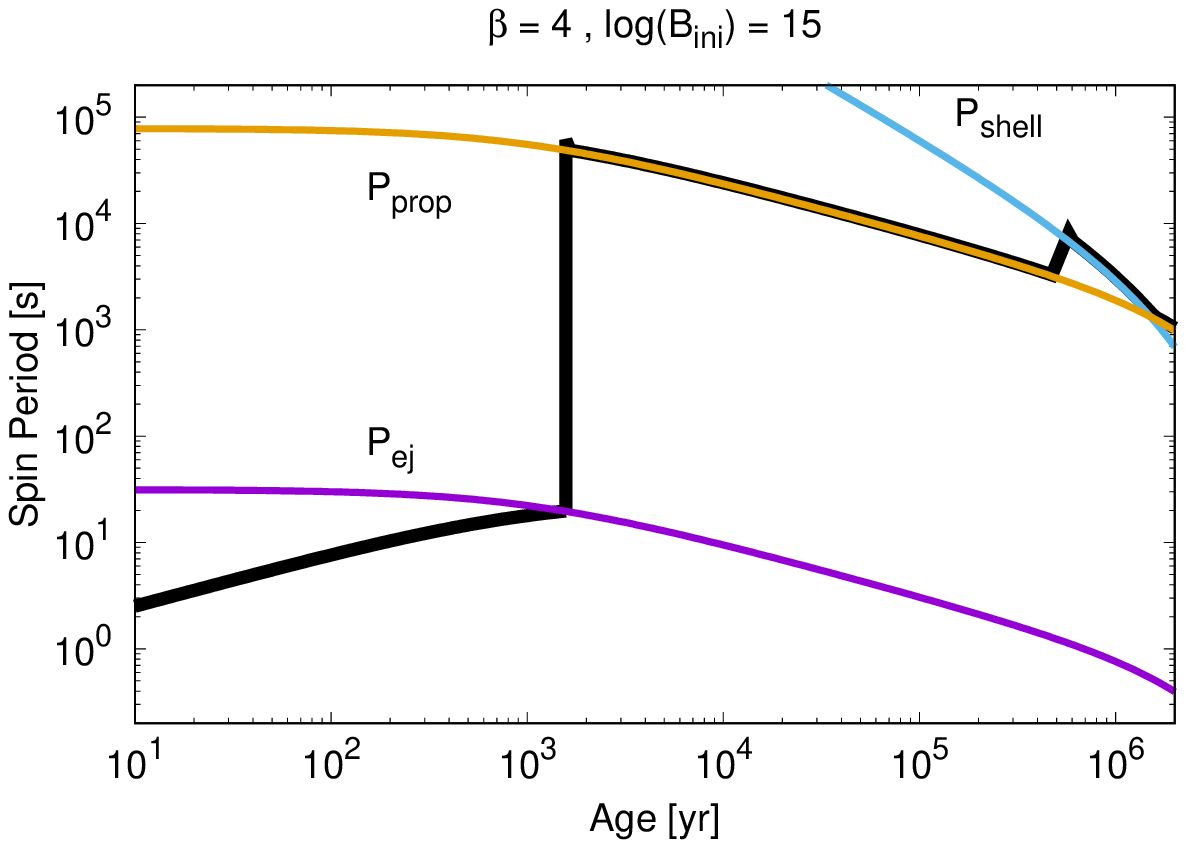} 
  \includegraphics[width=6cm]{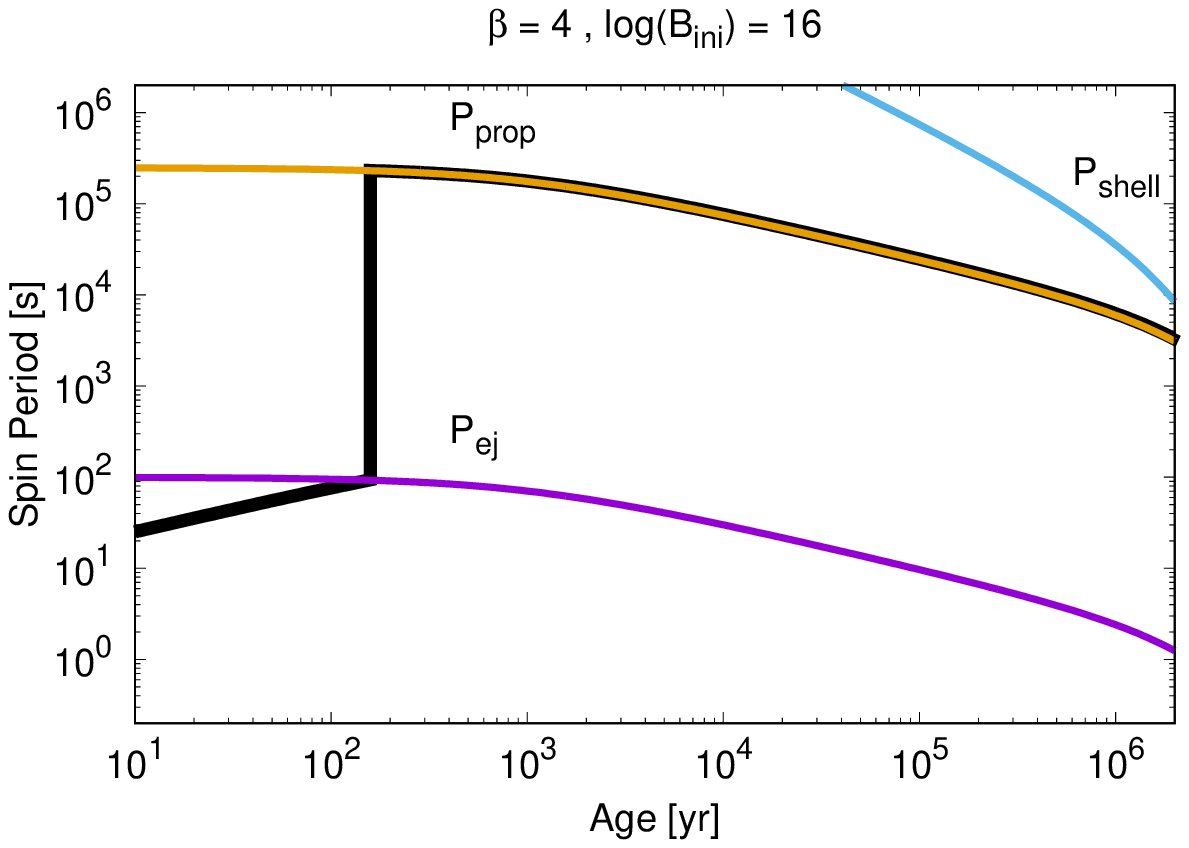}
\caption{The same figure with Fig.~\ref{fig:fast} but for intermediate wind velocity ($\beta = 4$).
}\label{fig:intermediate}
\end{figure*}

\begin{figure*}
  \includegraphics[width=6cm]{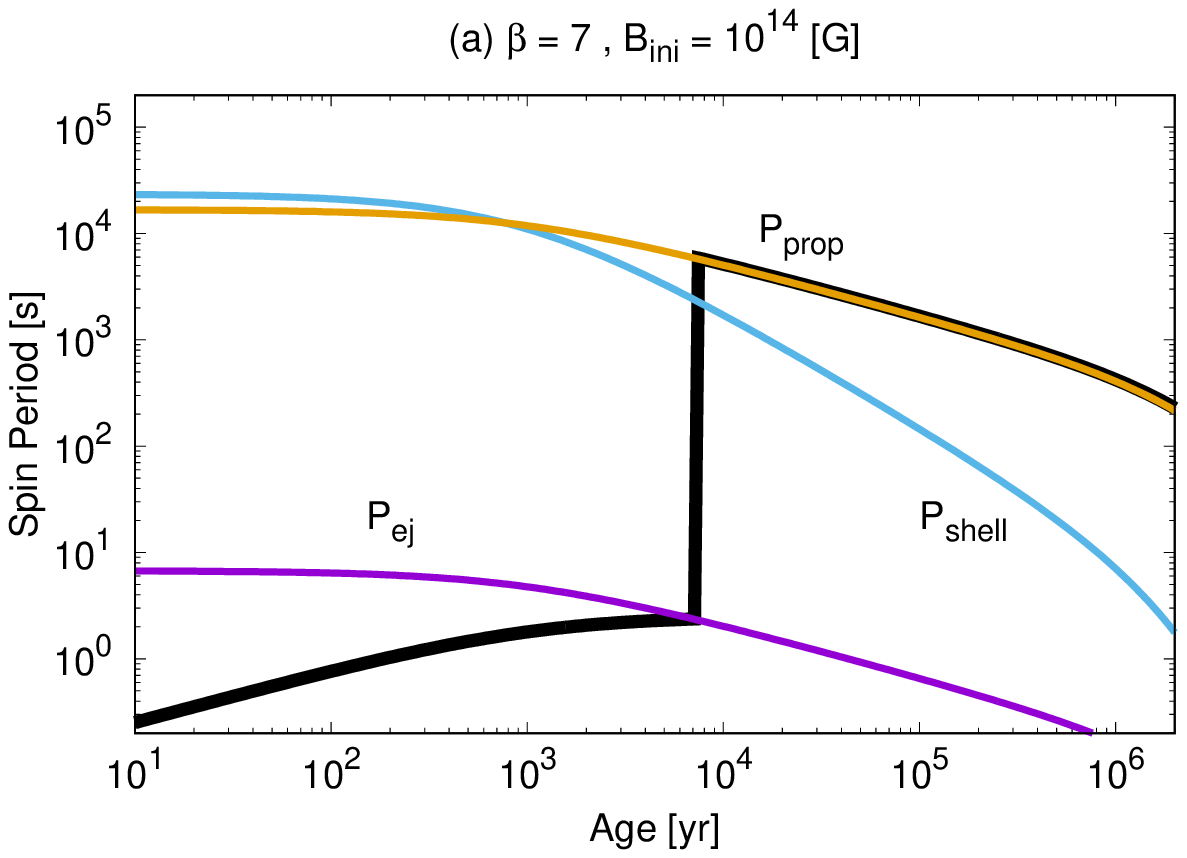} 
  \includegraphics[width=6cm]{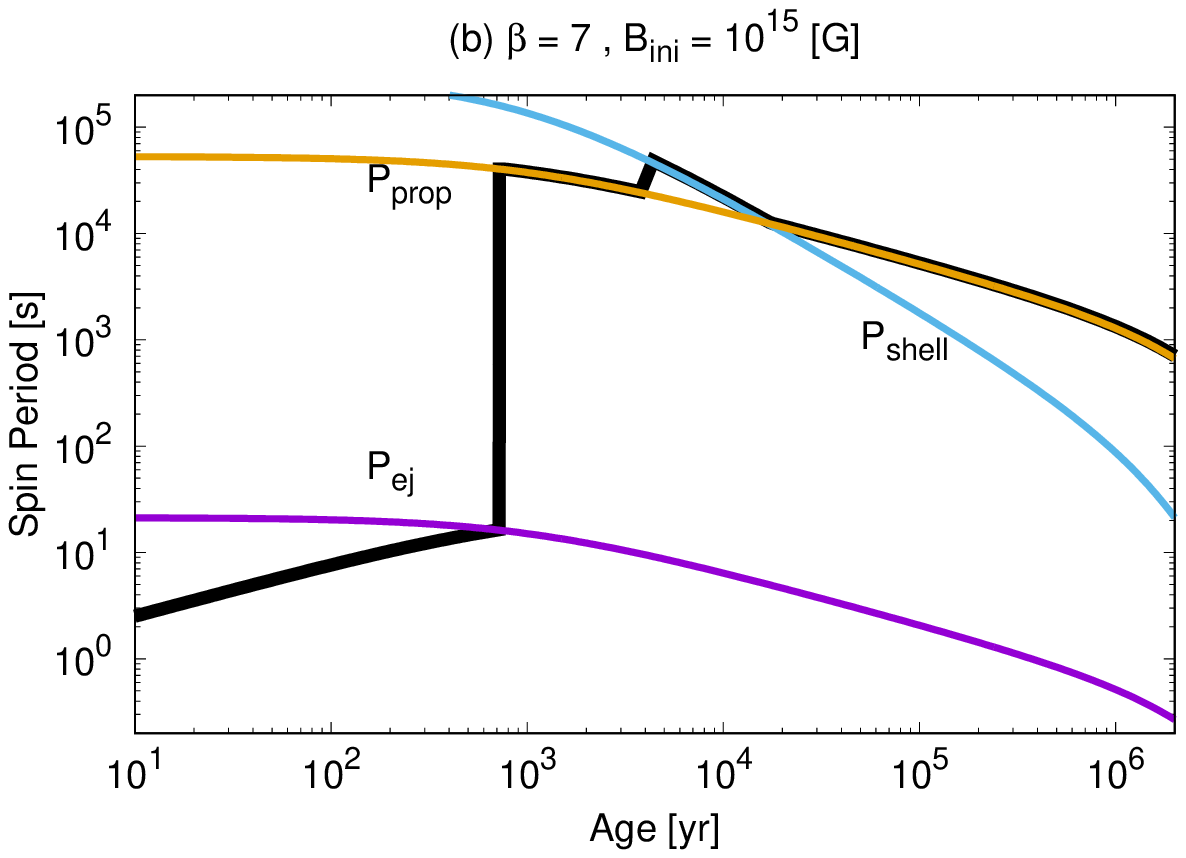} 
  \includegraphics[width=6cm]{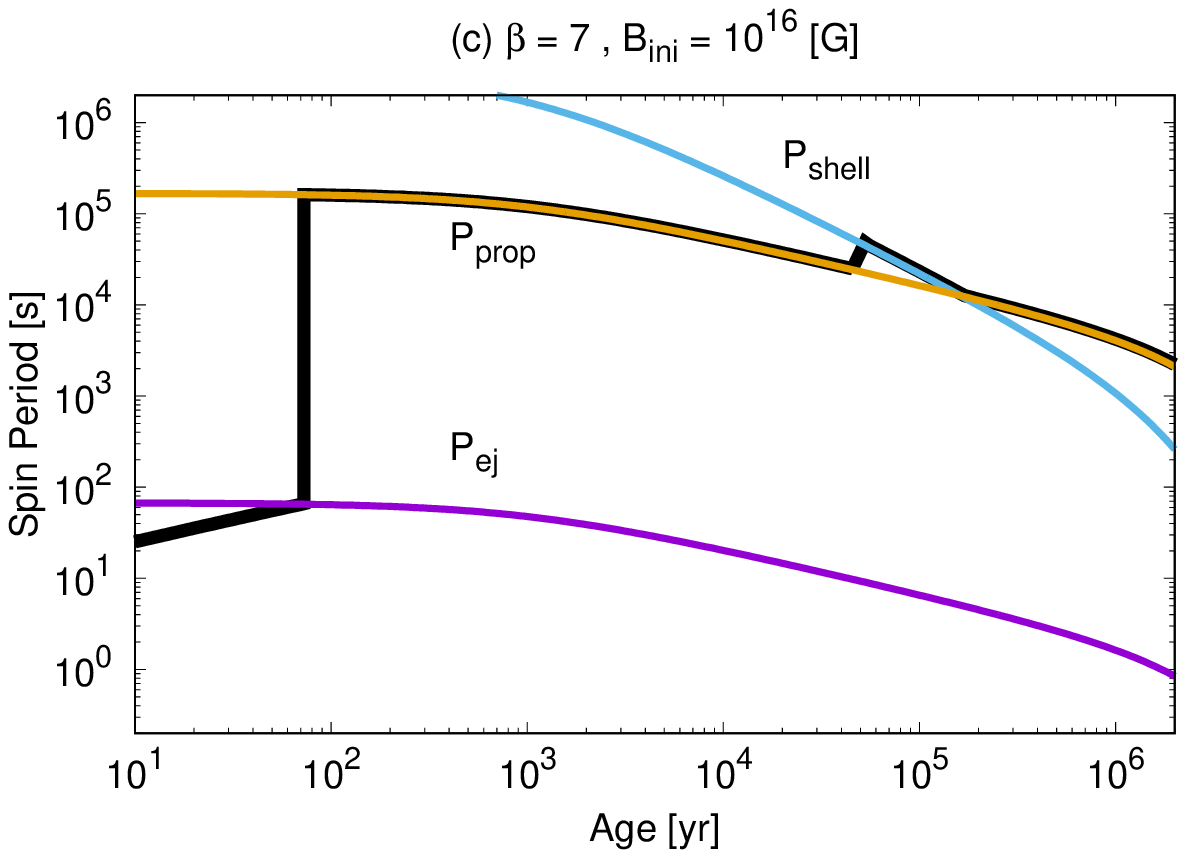}
\caption{
The same figure with Fig.~\ref{fig:fast} but for slow wind velocity ($\beta = 7$).
}\label{fig:slow}
\end{figure*}

\end{landscape}

Finally, we comment on the initial rotation of the neutron star. 
The faster (slower) rotation, the faster (slower) the spin-down in the ejector stage.
As a result, even if the initial spin is changed, it will asymptotically approach the $P_{\rm{spin}} = 10 \rm{ms}$ case in the early ejector stage, and then the same evolution is followed.

\section{Discussion}

The spin period of neutron stars in wind-fed systems show broad distribution spanning from $\sim 10$ to $\sim 10^{4} \, \rm{s}$ (see Table~\ref{tab:1}). 
Especially, recent monitoring observations revealed some slowly rotating neutron stars in HMXBs. 
Since it is considered that neutron stars are born with rapid rotation \citep{HLW00}, 
it is interesting question how to produce such slowly rotating neutron stars within short time after their birth \citep{LSL16,D16}.

As shown in the last section, we have performed model computations to investigate the spin evolution of slowly rotating X-ray pulsars.
Recently, \citet{D16} and \citet{WT20} applied propeller effect to examine the evolution scenario of the slowly rotation neutron stars in HMXBs. 
On the other hand, as the evolutional path of slowly rotating neutron stars, \citet{PT12} proposed an evolution scenario based on the settling shell formation.
In this scenario, after the end of ejector stage, the accretion shell is formed and dominates the spin of the neutron star. 
Since the spin-down due to the shell formation is so drastic, the neutron star spins down to $P_{\rm{spin}} > 10 \rm{ks}$ within short time-scale. 
Along this line, several works considering the evolution of neutron star spin have been tested by several authors \citep{Y19,S17,LSL16}.
The evolutionally track of such a system can be reconstructed in our model with moderately slow wind ($\beta \approx 5$) and weak magnetic field level.

In previous studies, however, only physical states after being bound by the neutron star gravity were taken into account implicitly. 
Hence, the magnetic inhibition, which is due to the interaction of the magnetic field with the unbound wind matter \citep{B08}, is omitted. 
Since the shell formation and propeller effect provides effective spin-down, the spin period of the neutron star easily reaches $> 10 \rm{ks}$. 
Actually, however, if the spin-down process due to magnetic inhibition is taken into account, the neutron star would spin down before the other processes occurs after the gravitational capture of the wind matter, as we have shown. 
By including the magnetic inhibition, it could be understood that most of neutron stars in wind-fed HMXBs settle down around the typical spin period of $1 – 10 \rm{ks}$ within their life-time ($\sim 1 \rm{Myr}$). 
In the last section, we have shown that the most effective physical process, among propeller effect, magnetic inhibition, and shell formation, depends on the stellar wind velocity.
We note that \citet{D16} argues that stellar wind velocity will not significantly affect spin evolution, but they arbitrarily gave the wind velocity and did not follow the time evolution of spin and the magnetic field, with stellar wind as a variable.
In fact, however, the stellar wind of the donor has a great influence on the physics of the neutron star spin evolution.

While the wind mass loss rate is taken from the well-tested recipe given by \citet{V01}, the wind velocity is a variable parameter in our study. 
It is broadly considered that in a HMXB system, the wind velocity of the donor is slower than that of isolated giant stars. 
This is because that the strong X-ray irradiation makes the donor envelope ionized and it reduces the efficiency of the line-acceleration \citep{S91,K12,K14}. 
Hence in most studies, relatively slow wind velocity is considered in the context of wind-fed HMXBs. 
On the other hand, if the X-ray luminosity of the neutron star is low, the ionization effect does not play important role, and consequently effective line-acceleration results in the fast wind velocity. 
Now that relatively dim HMXB is being observed with X-rays along with the improvement of X-ray observation, it is meaningful to consider the case of a fast wind accretion.

\begin{figure}
 \begin{center}
  \includegraphics[width=8cm]{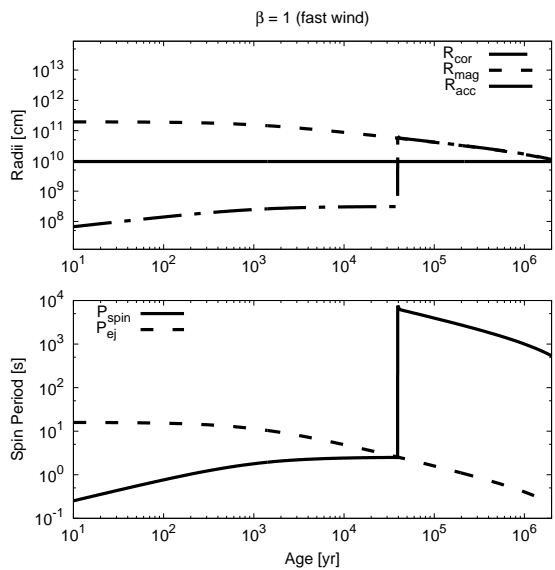} 
 \end{center}
\caption{
The evolutions of three radii: accretion radius $R_{\rm{acc}}$ (solid curve), 
magnetic radius $R_{\rm{mag}}$ (dashed curve) and corotation radius $R_{\rm{cor}}$ (dashed-dotted curve).
In the lower panel, the ejector condition $P_{\rm{ej}}$ (dashed curve )and the spin evolution of the neutron star 
(solid line) are also shown to see the end-point of the ejector phase. 
In the case with fast wind ($\beta = 1$), at the end of the ejector phase, since the magnetic radius is larger than 
the accretion radius, the magnetic inhibition dominates the spin evolution of the neutron star. 
The initial field is assumed to be $10^{14}\rm{G}$.
}\label{fig:PPP}
\end{figure}

\begin{figure}
 \begin{center}
  \includegraphics[width=8cm]{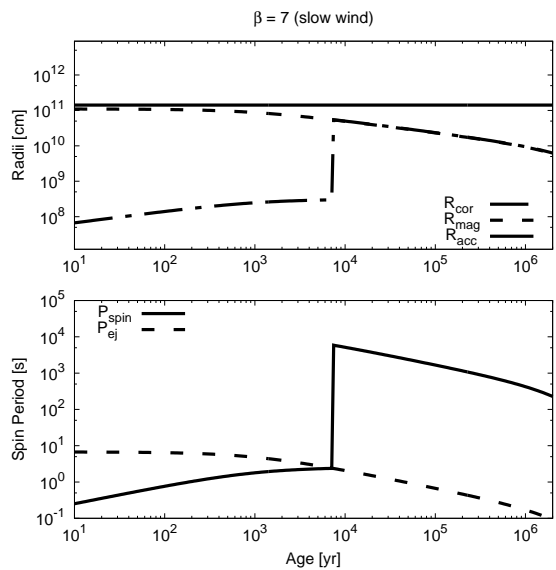} 
 \end{center}
\caption{
The same figure with Fig.~\ref{fig:PPP} but for slow wind velocity ($\beta = 7$).
In this case the accretion radius becomes larger than the magnetic radius, the classical propeller 
effect does work on the spin-down process of neutron star.
}\label{fig:QQQ}
\end{figure}

As discussed above, the spin down history after the ejector phase depends on the wind velocity. 
The dominant spin-down physics depends on the size relationship among $R_{\rm{mag}}$, $R_{\rm{acc}}$ and $R_{\rm{cor}}$. 
These size relations makes difference as seen in Figs.~\ref{fig:fast} -- \ref{fig:slow}.
In Figs.~\ref{fig:PPP} and \ref{fig:QQQ}, as a typical example, we show the size relationship between these three radii in the case of Figs.~\ref{fig:fast}(a) and \ref{fig:slow}(a).
In the upper panels of these figures, the evolution of three radii are shown, while in the lower panels the conditions of the ejector phase (see Eq.~(\ref{eq:eje})) are shown. 
When the wind velocity is slow, since the accretion radius becomes the largest just at the end of the ejector phase, the accretion shell formation dominates the spin evolution. 
On the other hand, when the wind velocity is fast, since the accretion radius is smaller than the magnetic radius, the spin is decelerated by the interaction between the magnetic field lines with the unbound wind matter. 
As seen in Figs.~\ref{fig:fast} -- \ref{fig:slow}, though the initial magnetic field affects the spin period, the wind velocity also affects the spin of the neutron star as well.

\citet{K12} has suggested that the velocity of the stellar wind in HMXB systems could be bimodal:
\begin{itemize}
\item {\it Fast wind and small X-ray luminosity}: 
When the wind velocity is fast, the mass accretion rate should be small in BHL theory and the consequent X-ray luminosity will be small. 
Hence, the ionization of the donor envelope becomes less effective. 
In this case, CAK mechanism works effectively and the wind velocity could be further faster. 
\item {\it Slow wind and large X-ray luminosity}:
When the wind velocity is small, the mass accretion rate should be large inversely. 
Hence, the X-ray luminosity will be large and the ionization of the donor envelope could be higher degree. 
In this case, CAK mechanism works less-effectively and the wind velocity could not be fast.
\end{itemize}
It has been shown that such kinds of bimodal solutions appear in one dimensional wind flow from massive donor to the compact object \citep{K14}. 
A fast wind accretion makes the accretion rate smaller, hence the X-ray luminosity from the neutron star should be small.
It means that opportunity to observe such fast wind systems should be small. 
This argument is consistent with the fact that the wind velocity of the donors in a certain fraction of the observed wind-fed systems are slower than isolated massive stars \citep{F15}.

Here, we note that in Eq.~(\ref{eq:Racc}), in practice, the relative velocity should be used rather than the wind velocity. 
In the present study, however, we want to focus on the effect of the stellar wind velocity, so we decided to use the wind velocity instead of the relative velocity. 
This is the same for Eqs. (\ref{eq:Pej}) and (\ref{eq:Rmag}).
In a binary system with a period of 5 days as we consider here, the orbital velocity cannot be completely ignored.
When the stellar wind is slow, the orbital velocity cannot be negligible with respect to the stellar wind velocity.
Therefore, we note that while the results for the slower stellar wind are qualitatively good, they may vary quantitatively.

\bigskip

As many authors discussed, slowly rotating neutron stars in HMXBs can be explained by rapid spin-down due to the propeller effect and/or the accretion shell formation \citep{PT12, D16,LSL16, S17, Y19,WT20}.
However, some neutron stars in HMXBs are rotating relatively rapidly, even though they are young. 
For example, well-studied HMXB system OAO1657-415 also shows short (38 seconds) pulse period \citep{M12}. 
Additionally, recently found HMXB system associated with a supernova remnant MC SNR J0513-6724 contains a neutron star which spin period is only 4.4 s \citep{M19}. 
The evolution path controlled by the propeller effect and/or shell forming cannot produce such a rapidly rotating neutron star, with standard level of magnetic field. 
If these objects are in spin equilibrium, the initial magnetic field of neutron stars should be $\sim 10^{11} \rm{G}$, and this value is extremely low for young neutron stars \citep{M19}. 
The spin evolution scenario of these rapidly rotating neutron stars in HMXBs is the next step to be tackled.

One speculation for understanding the spin evolution of rapidly rotating neutron stars in HMXB systems might be brought by different accretion geometry.
For example, the orbital period of MC SNR J0513-6724 is only two days, and the binary separation is quite small. 
In such a tight accreting system, it is not trivial that the wind matter is accreted onto the neutron star with keeping spherical symmetry. 
In asymmetric wind accretion, the shell formation scenario may be required some level of modification from the model originally proposed by \citet{S12}. 
Moreover, such a large asymmetric accretion may prefer disk formation. 
In fact, in several wind-fed systems, (temporal) formation of accretion disk is suggested \citep{HE13,N19,EM19b,T19,KNT19}.
Additionally, it is suggested that the focused wind accretion could make an effect on the mass and angular momentum transport in tight binary systems \citep{ HC12,EM19a}.

In the present study, we compute the spin evolution of the neutron star only within 1 Myr.
In fact, however, over 1 Myr, donor evolution cannot be ignored.
When the donor evolves, its radius expands and mass-loss rate becomes larger, and consequently the mass accretion rate onto the neutron star becomes larger. 
Once the mass accretion rate becomes larger than the critical value, the cooling time of the accreting matter becomes enough short to avoid forming settling shell \citep{S12}. 
In this case, the angular momentum associated with wind matter could be transported to the neutron star.
Then, the neutron star steps away from the quasi-equilibrium state and starts spin-up. 
The donor star in OAO1657-415 seems very large and could be evolved toward to the end of main sequence. 
If the donor star continues to evolve, the mass accretion rate becomes further large and the neutron star spin period will be few milliseconds. 
However, this phase could not be long since the evolution time-scale of massive star is very short.

When the surface of the donor is close to the Roche lobe, the accretion regime enters into the stage of wind Roche lobe over flow \citep{EM19a}.
According to the numerical calculations, the mass accretion rate is larger than the BHL model at this stage.
The spin of the neutron star will be further faster, but this stage has not been modeled yet in simple way.
The massive donor will fill its Roche lobe in short time scale and eventually absorb the neutron star into its envelope.
Such a system would become an ultra-compact binary or merge through common envelope evolution.
The common envelope evolution, however, has many unknown points such as the evaluation of recombination energy \citep{I13}.
This is a subject for further study,

\section{Conclusion}

In this study, we obtained the spin evolution models of neutron stars in wind-fed HMXBs. 
The neutron star stays in the ejector phase for a while after its birth, and then its spin period slows down until $P_{\rm{spin}} \sim 1 – 10 \, \rm{ks}$ rapidly. 
After the drastic spin-down, the neutron star spin settles down in quasi-equilibrium state and slowly spins up. 
In this study it is shown that the magnetic field evolution plays important roles on spin evolutions of the neutron stars in wind-fed binaries.
The magnetic field starts decaying about 1,000 years after the birth of the neutron star, causing a transition from the ejector stage to spin-down stage, and the subsequent spin-up also proceeds on the time scale of magnetic field dissipation.

Furthermore, it was shown that the dominant physical mechanism causing the spin-down depends on the accreting wind velocity. 
Roughly speaking, when the wind velocity is fast, the magnetic inhibition governs the spin evolution, while the shell formation and/or propeller effect becomes important in the slow wind case.
As we discuss, the wind velocity of the donor depends on the X-ray irradiation by the neutron star. 
Therefore, for the comprehensive investigation of the spin evolution of the neutron star in the binary system, in fact, it is necessary to trace not only the evolutions of the spin and magnetic field but also the evolution of the X-ray luminosity of the neutron star. 
Since X-ray luminosity is a major observable feature of HMXB along with the spin period, by solving these evolutionary paths consistently, the binary evolution process of HXMB could be correctly understood.
Since the evolution of HMXB also affects the population of pulsating ultra-luminous X-ray sources and the calculation of double neutron star merger rate, it could be an area of impact in astrophysics, and further research is expected in the future.

\begin{ack}
This work was supported by JSPS KAKENHI Grant Number 18K03706.

\end{ack}



\end{document}